\documentclass[journal,twocolumn,10pt]{IEEEtran} 
\usepackage{epsfig,cite}
\usepackage{tabularx}
\usepackage{multirow}
\usepackage{caption}
\usepackage{graphicx}
\graphicspath{ {./figures/} }
\usepackage{color}
\usepackage{amssymb,amsmath,mathrsfs}
\usepackage{float}

\newcommand{\rev}[1]{\textcolor{black}{#1}}

\begin{document}
\title{High-Speed Molecular Communication in Vacuum}
\author{Taha Sajjad and Andrew W. Eckford%
\thanks{The authors are with the Department of Electrical Engineering and Computer Science, York University, Toronto, Ontario, Canada M3J 1P3. Emails: tahaatiq@eecs.yorku.ca, aeckford@yorku.ca}%
\thanks{This work was supported in part by funding from the Lloyd's Register Foundation International Consortium of Nanotechnologies (LRF ICoN).}%
\thanks{Material in this paper was presented in part at the 2022 IEEE International Conference on Communications, Seoul, South Korea.}%
}

\maketitle
\begin{abstract}

Existing molecular communication systems, both theoretical and experimental, are characterized by low information rates. In this paper, inspired by time-of-flight mass spectrometry (TOFMS), we consider the design of a molecular communication system in which the channel is a vacuum and demonstrate that this method has the potential to increase achievable information rates by many orders of magnitude. We use modelling results from TOFMS to obtain arrival time distributions for accelerated ions and use them to analyze several species of ions, including hydrogen, nitrogen, argon, and benzene. We show that the achievable information rates can be increased using a velocity (Wien) filter, which reduces uncertainty in the velocity of the ions. Using a simplified communication model, we show that data rates well above $1 $ Gbit/s/molecule are achievable.
\end{abstract}



    


\section{Introduction}

Molecular communication promises to enable applications in important areas like nano-scale communication, and bioengineering \cite{farsad2016comprehensive}. A substantial body of research work has been done in this emerging area, and both theoretical \cite{infotheory,physicalmodel} and experimental \cite{Tabletop, barcodes}. Although the field is relatively new, molecular communication has many important applications, not only at the nanoscale but also in such applications as mines or pipelines where systems based on radio waves face significant path loss\cite{confined}.

To date, most work in molecular communication assumes the use of techniques such as diffusion, active transport \cite{onchip}, bacterial motility \cite{bacteria}, and diffusion with flow \cite{srinivas2012molecular,li2014capacity}. In terms of propagation environments, efforts have been made to analyze the information capacity of molecular communication in fluid\cite{Fluid} as well as in air \cite{Drift}, \cite{physical}.
These methods are biological or bio-inspired and result in very slow communication rates, even with advanced modulation techniques (e.g., MIMO \cite{MIMO} or acid-base signalling \cite{farsad2016molecular}); the best macroscale experiments have demonstrated on the order of ten bits per second. While such information rates are not an impediment in nature (e.g., in intercellular signalling transduction networks \cite{cheong2011information}), or in bio-inspired applications, where only simple control signals may be required, there is no reason in principle why information rates must be low in molecular communication.
Although the uncertainty inherent in diffusion-based communication arises from the very low mean free path of any individual molecule (leading to the Wiener process as a propagation model \cite{egan2021stochastic}),
this uncertainty can be counteracted by transmitting an enormous number of molecules, leading to an exceptionally high data rate \cite{eckford2014scaling}.

A different way to counteract the uncertainty and thus deliver high data rates for {\em small} numbers of molecules would be to find an environment in which the molecular velocities were relatively high while the mean free path was very long. At the microscopic level,  the velocities of the molecules (given by the Maxwell-Boltzmann distribution) are relatively high; while in high vacuum, the mean free path can be on the order of meters or kilometres \cite{chapman1990mathematical}. Thus, in a vacuum, it is possible to achieve data rates per molecule {\em many orders of magnitude higher} than have been demonstrated with existing systems. 

The main contribution of this work is to consider molecular communication in a vacuum as a potential candidate for a high-speed communication system. We illustrate the feasibility of this novel molecular communication method and show that high data rates are possible. Additionally, we improve the data rates further by using a {\em velocity filter}. The analysis is primarily inspired by time-of-flight mass spectrometry (TOFMS) systems, widely used in chemical analysis. However, a velocity selector is introduced to minimize uncertainty in the initial velocity of particles. Applications of our method may include communication in space, noting that particle-based schemes have previously been shown to be an energy-efficient method for space communication \cite{Inscribedmatter,commInscribed}.  However, the process can also be applied to terrestrial systems using vacuum tubes; given the achievable rates, this method presents a promising alternative to conventional communication methods. 



The paper is organized as follows.  In Section II, we describe our basic system and its theoretical aspects. In Section III,  we present a communication model to analyze the performance of our proposed system. In Section IV, we will discuss results in terms of data rate. Section V will focus on the conclusion and future prospects in this domain.





\section{System model}

\subsection{Overview of physical model}

Here we review the physical basis for our system, which is inspired by the operation of time-of-flight mass spectrometry (TOFMS) \cite{ResolutioninlinearTOF, PrinciplesTOF, ImprovedResolution}. (Throughout this paper, the acronym TOFMS will refer to both the concept of {\em spectrometry} or the {\em spectrometer} device.) A basic TOFMS consists of an {\em ion source} at the transmitter end, a {\em collector} on the receiver end, and a {\em drift region} in a vacuum that connects the transmitter to the receiver. 
%
%
 \textcolor{black}{Ions removed from a sample of an element may be accelerated by an electric field source, which assists them to flow towards the receiver}. 

\begin{figure}[t!]
\centering
\includegraphics[width=1\columnwidth]{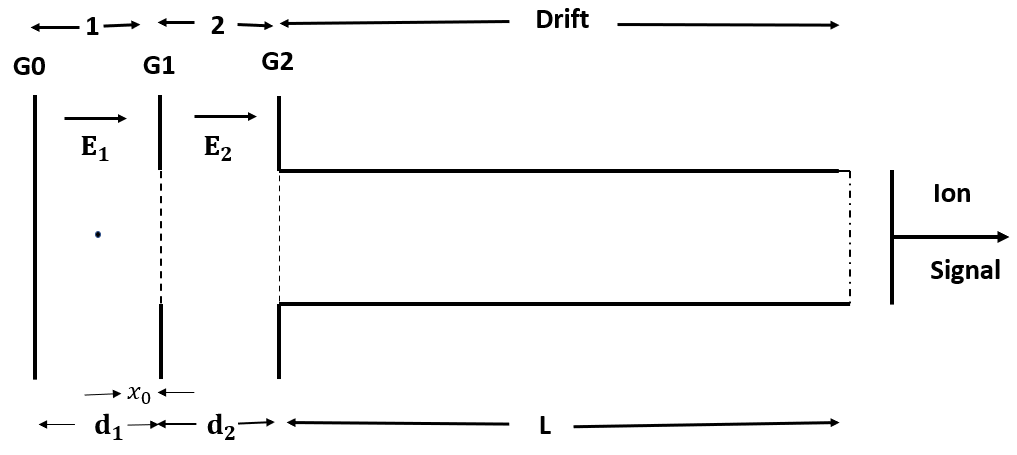}
\caption[\hspace{0.2cm}]{Transmitter consists of two regions 1 and 2. Two accelerating electric fields are generated by applying voltages $V_0$, $V_1$ and $V_2$ between Grids $G0$, $G1$ and $G2$, respectively. The distance between the ionization point and grid $G1$ is $x_0$, the distance between grids $G0$ and $G1$ is $d_1$, the distance from $G1$ to $G2$ is $d_2$, and the length of the drift tube is $L$.
Figure adapted from \cite{ResolutioninlinearTOF}.}
\label{fig1}
\end{figure}

We follow the setup in \cite{ResolutioninlinearTOF} to analyze the motion of ions in a vacuum under the influence of an electric field, taking into account the effects of initial velocity, spatial spread, and temporal profile of the ionization source. This system is shown in Fig. \ref{fig1} from \cite{ResolutioninlinearTOF}, which shows a double-field source containing two accelerating regions instead of a single-field source; the purpose of using the double source is to reduce the distribution by reducing initial velocity and spatial spread \cite{ImprovedResolution}, which is useful both in TOFMS and communication.

In the double-field source, potentials $V_0$, $V_1$ and $V_2$ are applied to grids $G0$, $G1$ and $G2$, as depicted in Fig. \ref{fig1}. In the model, a single atom or molecule with an initial velocity component $v_{0}$ is ionized at a distance $x_{0}$ from grid $G1$, forming an ion of mass $m$ and charge $q$. The ion experiences an electric field $E_{1}$ in the first region and is accelerated towards the second region with an electric field $E_{2}$ before emerging into the drift region.
This ion is the information carrier in our system. 

 The amount of time taken for ions to traverse the drift region is of key importance to our analysis. Ideally, if an ion's flight time depends only on its mass-to-charge ratio, then TOFMS should have an unlimited resolution. However, this is not the case. In practice, an ion's flight time depends on many different factors besides mass-to-charge ratio, including space charge effects, inhomogeneous electric field, the finite frequency response of detector, spatial and temporal spread and initial distribution of ion velocities \cite{ResolutioninlinearTOF}, and many methods have been applied to improve these factors \cite{PrinciplesTOF, ResolutioninlinearTOF}. Our modelling, as well as in \cite{ResolutioninlinearTOF}, is focused on variations influenced by the initial thermal velocity and spatial spread. Mathematical expressions will be derived in the next section.




 \subsection{Time-of-flight analysis}
 
 Here, following \cite{ResolutioninlinearTOF}, we obtain a distribution for the time-of-flight of an ion launched from a double-field source through a vacuum drift region of length $L$. 
 
 We define the $x$ direction as the direction from transmitter to receiver, and we assume that only the velocity component in this direction is relevant. Equivalently, we assume that the system is uniform in the $y,z$ plane; that is, there is no change in the $y,z$ directions. It is a common assumption when analyzing molecular timing channels \cite{nakano2013molecular}.
 
 In the system depicted in Fig. \ref{fig1}, an electric potential is applied to the ion, and potential energy gained by ions in the accelerating region is converted to kinetic energy. From basic electrostatics,
 \begin{align}
   \label{eqn:drifttime}
   qV &= \frac{1}{2}mv^2\, ,
 \end{align}
However, since $v = L/t$\,,
\begin{align}
   t &= L\sqrt{\frac{m}{2qV}}\, .
\end{align}
 Here $V$ is the total voltage from Grid $G0$ to $G2$ while $v$ is velocity. Thus heavier ions tend to reach later than lighter ones. The ions are formed in the ionization region of the source with some initial velocity distribution. Under normal experimental conditions, this  is Maxwell-Boltzman distribution which is a one-dimensional distribution with zero mean and with standard deviation as $\sqrt{kT/m}$, where $T$ is temperature and $k$ is Boltzman constant. These newly generated ions are all accelerated towards the detector by one or a series of constant electric fields\cite{ImprovedResolution}.

  Following \cite{ResolutioninlinearTOF}, the probability density function PDF used to describe the ion's profile peak is Gaussian, where the $x$-axis represents time $\tau$, and the $y$-axis represents probability density function $f(\tau)$. The probability of an ion having flight time in a specific interval can be calculated by integrating the flight time PDF over that range. We can describe the initial velocity and spatial distribution of the laser ionization source as a Gaussian distribution with the random variable $v_{0}$, $x_{0}$ and $t_{0}$ \rev{where $T(x_{0},v_{0})$ as the average time- of-flight starting from position $x_{0}$ and with initial velocity $v_{0}$.} The joint PDF of  $v_{0}$ and $x_{0}$ and $\tau$ is 
  \begin{align}
   \nonumber\lefteqn{f(x_{0},v_{0},\tau) =} &\\ 
    \label{eqn:jointpdf}
   & \:\:\:\: Ce^{-[(\tau-T(x_{0},v_{0}))^2/2\sigma_{t}^2]+[(x-x_{0})^2/2\sigma_{x}^2]+[(v-v_{0})^2/2\sigma_{v}^2]}.
\end{align}
  Here $\sigma_{t},\sigma_{x},\sigma_{v}$ are the standard deviation of time, position and velocity. In \cite{ResolutioninlinearTOF}, standard deviations have been calculated from the full width at half maximum 
  of the laser's beam diameter and its temporal duration. 
  Moreover, $C$ is the normalization constant and depends upon above mentioned standard deviations:
 \begin{align}
   \label{eqn:c}
   C &= \frac{1}{(2\pi)^{3/2}\sigma_{t}\sigma_{x}\sigma_{v}}\, .
\end{align}

  We are interested in the marginal distribution of time $\tau$, so we integrate both position and time to obtain
  \begin{align}
   \label{eqn:marginal pdf}
   f(\tau) &= 
   \iint f(x_0,v_0,\tau)\,dx_{0}\,dv_{0}\,.
\end{align}

 We adapt the results from \cite{ImprovedResolution} in order to calculate $T(x_{0},v_{0})$. However, these results were calculated in terms of kinetic energy instead of the initial velocity. Therefore, we will first describe the set of equations mentioned in \cite{ImprovedResolution} before expressing them in terms of the initial velocity. We use the following relations between velocity $v$ and kinetic energy $U$:
 \begin{align}
  \label{eqn:Total Time}
    U_{0} &= 1/2 mv_{0}^2\, , \\
    U     &= 1/2 mv^2\, , \\
    &= U_{0}+ qx_{0}E_{1} + qd_{2}E_{2}\, , \\
    v &= \sqrt{v_{0}^2+\frac{2}{m}qx_{0}E_{1}+\frac{2}{m}qd_{2}E_{2}}\, .
\end{align}

Here $v$ and $U$ are total velocity and kinetic energy, respectively; $v_{0}$ and $U_0$ are initial velocity and kinetic energy, respectively. $E_{1}$ is an electric field between $G0$ and $G1$ while $E_{2}$ is an electric field between $G1$ and $G2$. Under these conditions, the time-of-flight has been given in \cite{ImprovedResolution} as

 \begin{align}
   T(U_{0},x_{0}) &= T_{x_{0}}+T_{d2}+T_{L}\, ,
 \end{align}
 where
 \begin{align}
   T_{x_{0}} &= 1.02 \left[\frac{\sqrt{2m[U_{0}+qx_{0}E_{1}]}}{q_{1}E_{1}}\pm \frac{\sqrt{U_{0}}}{qE_{1}}\right]\,  , \\
   T_{d_{2}} &= 1.02\frac{\sqrt{2m}}{qE_{2}} \left[\sqrt{U}-(\sqrt{U_{0}+qx_{0}E_{1}})\right]\, , \\
   T_{L} &= 1.02\frac{\sqrt{2m}L}{2\sqrt{U}}\, .
 \end{align}
 From (6) and (9), we can easily express (10) to (13) in terms of initial velocity as
 \begin{align}
   T(x_{0},v_{0}) &= T_{x_{0}}+T_{d2}+T_{L}\, ,
 \end{align}
 where
 \begin{align}
   T_{x_{0}} &=\frac{1.02m}{qE_{1}}\left[\sqrt{[v_{0}^2+\frac{2}{m}qx_{0}E_{1}]}\pm v_{0}\right]\, , \\ 
   T_{d_{2}} &= 1.02\frac{m}{qE_{2}}\left[v-\Bigl(\sqrt{v_{0}^2+ \frac{2}{m}qx_{0}E_{1}}\Bigr)\right]\, , \\
   T_{L} &= 1.02\frac{L}{v}\,.
 \end{align}
 In (\ref{eqn:Total Time}) $U_{0}$ is the initial kinetic energy which is derived from the initial velocity in our work. $U$ is total energy, including initial kinetic energy and energy gained by ion while passing through the field regions 1 and 2. Equ. (9) calculates the time of flight required by an ion. A detailed derivation of this set of equations is given in \cite{PrinciplesTOF} and \cite{ImprovedResolution}. There are three types of time delays mentioned in the literature. $T_{x_{0}}$ corresponds to the time an ion takes when it is produced from the source to reach grid $G1$. The second term in the same equation represents turnaround time. If there is an ion whose direction of movement is opposite to the grid, it requires time to turn around toward grid $G1$. Hence $\pm$ corresponds to the initial velocity directed towards or away from the detector. $T_{d_{2}}$ is the time required to pass through distance $d_{2}$, and finally, $T_{L}$ is the travelling time of an ion to pass through in a drift region. We can modify our basic system by introducing a velocity selector known as the Wien filter with the hope of reducing uncertainty in the time-of-flight and hence improving data rates.  
 
 \begin{figure}[ht!]
\centering
\includegraphics[width=\columnwidth]{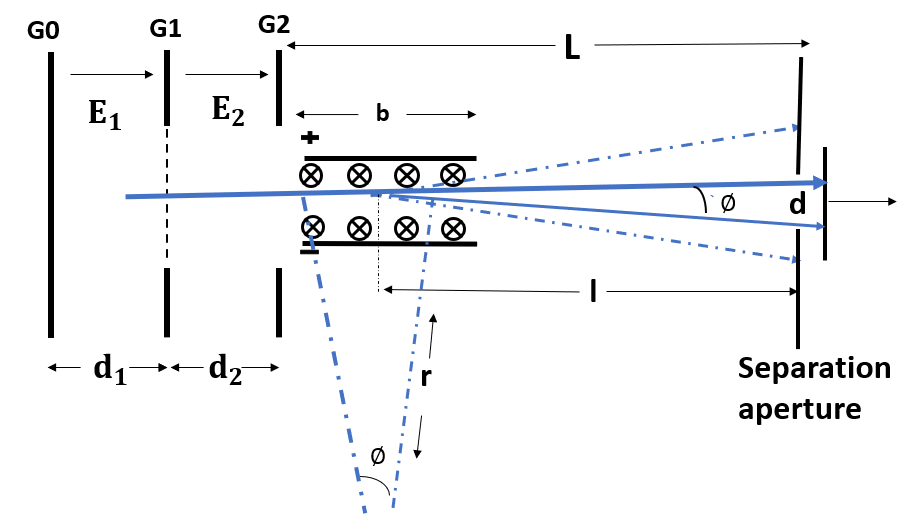}
\caption[\hspace{0.2cm}]{Transmitter consists of two regions 1 and 2. Two accelerating electric fields are generated by applying voltages $V_0$, $V_1$ and $V_2$ between Grids $G0$, $G1$ and $G2$, respectively. A velocity selector is added, having a positive charge in the upper plate and a negative charge in the lower plate. While the magnetic field has direction inside of the paper. Electric and magnetic fields are arranged in such a way that will allow ions of specific  velocity to pass straight from the device while other ion beams bend in different directions. The selector length is $b$, whereas $l$ is the length from the centre of the device to the separation aperture of size $d$.The distance between the ionization point and grid $G1$ is $x_0$, the distance between grids $G0$ and $G1$ is $d_1$, the distance from $G1$ to $G2$ is $d_2$, and the length of the drift tube is $L$.
Figure adapted from \cite{ResolutioninlinearTOF}.}
\label{fig2}
\end{figure}

\subsection{Velocity Selector (Wien filter)}
A velocity selector, also known as a Wien filter, is a device that employs magnetic and electric fields to select charged particles of a particular velocity \cite{schmidt2009compact}, which has been used in a chemical analysis \cite{seliger1972b}, \cite{srinivas2012molecular},\cite{ap1997determination}. Fig. \ref{fig2} shows the perpendicular arrangement of the electric and magnetic field required in principle to allow a particle of a particular velocity to emerge. You can find the fundamental relationship between selected velocity $v_s$ and electric field $E$ and $B$ in any related reference, for example, \cite{seliger1972b} and \cite{schmidt2009compact}. A charged particle experiences a Coulomb force while moving in an electric field $E$.
\begin{align}
  \label{eqn:particle in electric field}
    F_{c} &= q E\,, 
\end{align}
 parallel in the direction of the electric field. However, a uniform magnetic field tends to move a particle in a circular path under the Lorentz force given by
\begin{align}
  \label{eqn:particle in magnetic field}
    F_{m} &= q\vec{v_s} \times \vec{B}\,, \\
          &= qv_sB\sin{\theta}\,.
\end{align}
Here $v_s$ is a selected velocity. In our setup, the beam of ions enters a velocity selector at an angle of $90$ degrees, experiences a downward electric force, and is driven upward by a magnetic force. Since $\sin{\theta}= \sin{90}=1$, the net force on ions is 
\begin{align}
  \label{eqn:net force}
    F_{m} &= qv_sB\,, \\      
    F     &= qv_sB - qE\,,
\end{align}

 It is essential that both (magnetic and electric) forces balance so that an ion can move undeflected. Ions with velocity $v_s$ passing through the velocity filter won't deflect if $F= 0$. 
\begin{align}
        0  &= qv_sB - qE\,,\\
        v_s  &= E/B\,.
\end{align}
In the expression above, only a selected component of velocity $v_s$ should pass through the aperture. However, the size of the aperture $d$ can be detrimental to accuracy and a source of uncertainty. Here we will derive a general relationship between velocity $v$ and aperture $d$ based on \cite{waahlin1964colutron}. 
Whereas velocity $v$ of ion accelerated through electric potential across grids from \rev{(\ref{eqn:drifttime})} is
\begin{align}
        v &= \sqrt{\frac{2qV}{m}}\,,
\end{align}
We can express the relationship between electric and magnetic fields using (24) and (25) as
\begin{align}
        E &= B\sqrt{\frac{2qV}{m}}\,,
\end{align}
Ions moving with velocities $v$ other than our selected velocity $v_s$ experience centripetal force. Therefore, (23) becomes
\begin{align}
    F     &= qvB - qE\,,\\
    \frac{mv^2}{r} &= qvB - qE\,, 
\end{align}
From (27), (30) becomes
\begin{align}
    \frac{2qV}{r} &= qvB - qE\,,
\end{align}
Combining (26) and (31)
\begin{align}
    \frac{2V}{r} &= \frac{vE}{v_s} - E\,, \\
              r  &= \frac{2V}{E(\frac{v}{v_s}-1)}\,.    
\end{align}
Here $r$ is the radius of the circular path around which the velocity components, slightly different from the exact value of $v_s$, tend to revolve. The size of  the aperture determines the degree of velocity variations which can pass other than selected velocities.

From Fig. \ref{fig2}, $l$ is the distance from the centre of velocity selector to the aperture, $b$ is the length of the filter, $d$ is aperture size, and $\phi$ is the angle between selected velocity $v_{s}$ and other velocities which can pass through the aperture. Since
\begin{align}
 \label{eqn:aperture}
    d &= l\phi\,,
\end{align}
and
\begin{align}
    \phi &= \frac{b}{r}\, ,
\end{align}
we can write
\begin{align}
    d &= \frac{lb}{r}\,,
\end{align}
From (33)
\begin{align}
\label{eqn:velocityaperture}
    d &= \frac{lbE}{2V}\frac{(v_s - v)}{v}\,.
\end{align}
The size of the aperture is directly related to $\delta v$ (the range of velocity that can pass through the aperture) along with other parameters. By decreasing the aperture size infinitesimally small, we can avoid {\em unwanted} components of velocity, but there is a chance that the {\em selected} component may be {\em missed}.

\section{Communication model}

Our goal is to demonstrate the feasibility of molecular communication in a vacuum and to demonstrate that this type of communication is capable of extremely high data rates. To this end, we make simplifying assumptions in our communication model for the system with or without a velocity selector so as to highlight the rates that are achievable in principle with this form of communication.

\subsection{Communication system components}

In Fig. \ref{fig1}, the communication model is overlaid with the physical schematic, depicting the source, drift region, and collector as a transmitter, channel, and receiver, respectively. The components of this system are described as follows.
 %
%

\begin{itemize}

    \item {\em Transmitter.}
     The transmitter consists of the ionization source and controlling mechanism to generate and release ions in a controlled manner. Many techniques might be used to ionize molecules; in work referred to in \cite{ResolutioninlinearTOF}, a Lumonics TE-430 XeCl excimer laser is used. A control mechanism is required to encode a message according to the modulation scheme, described in more detail below. We assume that the transmitter can control the release and ionization of molecules precisely, but cannot control the initial velocity and initial location of the ion. From the transmitter, the ions acquire high velocity while passing through grids G1 and G2 in the direction of the receiver.
     
    \item {\em Channel.}
     The drift region forms the channel. It is assumed that the drift region is evacuated to a high vacuum, and there are no traces of any other gas (i.e., the mean free path is much longer for the ions). 
     
    \item {\em Receiver.}
     Ions arriving from the drift region are collected by the receiver, which could be a simple metal detector, an electron multiplier, a counter, or time to digital converter; examples are described in \cite{Detector}. (In \cite{ResolutioninlinearTOF}, Varian microchannel plates were used as a detector.) Whatever the mechanism, the receiver records the arrival time of the ion. For the purposes of this analysis, we assume that the arrival time is recorded precisely, without noise or error.
     
     \item {\em Velocity Selector.}
     We follow the setup in \cite{waahlin1964colutron} to analyze the impact of the velocity filter on data rates. The velocity filter in \cite{waahlin1964colutron} is composed of an electromagnet with two electrostatic deflection plates mounted between the pole tips. The magnet weighs approximately 200 kg, and it has the ability to develop a maximum magnetic field of $12000$ Gauss. The electrostatic deflection plates have the same length as that of the magnet pole tips. In order to ensure a uniform electric field between the deflection plates, a set of 18 guard rings is provided with adjustable potentials.

\end{itemize}
In the following subsection, we will discuss modulation and detection schemes for both of our proposed systems separately and will analyze the  sources of uncertainties using information theoretical aspects.

    
\subsection{Modulation schemes and detection for basic system}


From this description, it should be clear that the main source of uncertainty in the system is the random initial velocity and location of the particle. 
For simplicity, we consider on-off keying modulation: time is discretized into intervals of length $T$. A single ion is released for bit 1, and no ions are released for bit 0.
The transmission may be represented as a sequence of transmitted bits $\vec{b} = [b_1,b_2,\ldots]$, where $b_i \in \{0,1\}$ is the bit transmitted in the $i$th interval. 

Equivalently, since it will be convenient to represent the channel as a timing channel, we can represent $\vec{b}$ as a vector of release times.
Suppose the transmitter releases $m$ molecules. The release times of the molecules can be represented by $\Vec{x}=[x_1,x_2,...,x_m]$; for example, if $b_i = 1$ is the $j$th molecule to be released, then $x_j = i T$. The particle released at time $x_i$ arrives at time $x_i+t_i$. Furthermore, the propagation times of the molecules can be represented by $\Vec{t}=[t_1,t_2,...,t_m]$. Then the receiver observes the vector of arrivals form of vector $\vec{y}=\vec{x}+\vec{t}$ where $\Vec{y}$ is the first arrival time of particles at the receiver. We interpret $\vec[t]$ as additive noise, which is uncertain due to the random initial velocity.

Now consider the distribution of the noise $\vec{t}$. Assuming molecules are distinguishable so that output $y_i$ corresponds to input $x_i$, the probability density function of $y_i$ given $x_i$ can be written as 
\begin{align}
    f_{Y_i|X_i}(y_i|x_i) = f_t(y_i-x_i) , 
\end{align}
with first arrival time given by (\ref{eqn:marginal pdf}).

\textcolor{black}{In terms of detection, we attempt to detect bit $i$ only in the interval $(iT,(i+1)T)$. Moreover, we make the simplifying assumption that {\em if a molecule is emitted at time $i$, and does not arrive by time $(i+1)T$, it is removed from the system}.} Thus, the probability of {\em not} arriving during this interval, and hence the probability of missed detection (observing 0 when 1 was transmitted), is given by
\begin{align}
    \label{eqn:probofmisseddetection}
    P_d &= 1 - \int_{0}^{T} f_t(t) dt .
\end{align}
Under our assumptions, note that the probability of a false alarm (observing 1 when 0 was transmitted) is zero. For the purpose of demonstrating the feasibility of molecular communication in a vacuum, the assumption of removing molecules is reasonable if $P_d$ is small. Moreover, it simplifies the problem to finding the interval $T$ that reduces $P_d$ to a sufficiently small value; the achievable data rate is then
\begin{align}
    \label{eqn:DataRate}
    R = \frac{1}{T}\,. 
\end{align}
in bits per second.
\textcolor{black}{In this paper, we obtain $T$ corresponding to $6$ standard deviation surrounding the mean (i.e., $\pm 3$ standard deviations from the mean), which implies $P_d \leq 0.01$. Combined with our simplifying assumption that ``late'' molecules are removed, this gives an upper bound on the information rate, but one that is close to the actual achievable rate.}

While a more accurate achievable rate could be obtained from a high-fidelity model, this simplified rate is sufficient to demonstrate the feasibility of the scheme. 



\subsection{Modulation and Detection Scheme for a system with Wien Filter}
As mentioned in the previous section, uncertainty lies mainly in the randomness of the initial velocity of the particles. The purpose of using a Wien filter is to minimize this uncertainty by ensuring that a beam of ions with a specific velocity will pass. The question arises, what is the uncertainty in the ultimate velocity? Moreover, what uncertainty is introduced by discarding molecules that do not pass through the filter? This section addresses these questions using an information-theoretic analysis of the channel, which will help us determine the true achievable rate. We will use the same approach as used in \cite{chouhan2019optimal} which is finding out the probabilities of correct and false messages and then using mutual information to determine the rates.

Consider an on-off keying modulation: a beam of ions with a specific velocity $v_s$ is released as bit 1 whereas no ions are released for bit 0. According to (\ref{eqn:velocityaperture}), a range of velocities that can pass through the aperture depends on the aperture size, let's suppose $v_s \pm\delta v$ is a range of velocity that comes out of the Wien filter. Then the probability of this certain range can be found out as

\begin{align}
    \label{eqn:pv}
    p_v &= \int_{v_{s}-\delta v}^{v_{s}+\delta v} f(v) dv.
\end{align}
where $f(v)$ is the Maxwell-Boltzmann distribution. By the time $t = T$, the receiver has either observed the beam of ion arrives ($y = 1$) or not ($y = 0$). Thus, we can write the conditional probability
\begin{align}
    \label{eqn:likelihood}
    p(y|x) &= 
    \left\{
        \begin{array}{cl}
             1, & y=0,x=0; \\
             0, & y=1,x=0; \\
             p_v, & y=1,x=1; \\
             1-p_v, & y=0,x=1.
        \end{array}
    \right.
\end{align}
Hence mutual information is given by 
\begin{align}
    \label{eqn:mutual}
    I(X;Y)= H(Y)-H(Y|X)\,,
\end{align}
We now use the channel model to calculate the capacity.
\rev{First we calculate the probability $p_y(y)$, as follows:
  \begin{align}
      p_{Y}(y) &= \sum_{x\in\{0,1\}}p_{Y|X}(y|x)p_{X}(x)\,,\\
           &= 
           \left\{ \begin{array}{cl} 
           p_{X}(1)(1-p_v) + p_{X}(0), & y=0\,, \\
           p_{X}(1) p_v, & y=1\,,           
           \end{array} \right. 
   \end{align}
from which $H(Y)$ can be obtained as
\begin{align}
    H(Y) = \sum_{y \in \{0,1\}} p_Y(y) \log \frac{1}{p_Y(y)} .
\end{align}
}
Subsequently, we obtain $H(Y|X)$ from 
  \begin{align}
   H(Y|X)= \sum_{x\in\{0,1\}}\sum_{y\in\{0,1\}} p_{X,Y}(x,y)\mathrm{log_2}\frac{1}{p_{Y|X}(y|x)}\\
         = \sum_{x\in\{0,1\}}p(x)\sum_{y\in\{0,1\}}\mathrm{log_2}\frac{1}{p_{Y|X}(y|x)}\,.
  \end{align}
   \rev{Let
  \begin{align}
    \mathcal{H}(p_v)= p_{v}\mathrm{log_2}\frac{1}{p_{v}}+(1-p_{v})\mathrm{log_2}\frac{1}{1-p_{v}} 
  \end{align}
  represent the binary entropy function. Then from (\ref{eqn:likelihood})
\begin{align}
  \label{eqn:binary entropy}
      H(Y|X)&=p_{X}(1)\mathcal{H}(p_v) + p_X(0)\mathcal{H}(1) \\
      &= p_{X}(1)\mathcal{H}(p_v) .
  \end{align}}

   Since mutual information is given by (43), the Information rate can be calculated as
   \begin{align}
      \label{eqn:infotheoretic rates}
      R &= \frac{I(X;Y)}{T}\,.
   \end{align}
   Here $T$ is calculated in the same way as in (17), except that $v$ is replaced with $v_s$. In the next section, we will analyze the results obtained from our proposed systems.
\section{Results and Discussion}

Here we present results illustrating the feasibility of molecular communication in a vacuum. As the velocity distribution depends on the mass of the emitted ion, we consider atoms of hydrogen, nitrogen, and argon. Although hydrogen is highly flammable, it was chosen for illustration because it has the lowest mass, making it faster. Nitrogen is abundant in the atmosphere, while argon is an inert gas.  We also consider ionized molecules of benzene, which is a common molecule for this kind of analysis in the TOFMS literature. Table I shows the parameters used in a simulation for the analysis of communication systems.

\subsection{Analysis of arrival time distributions}
\begin{figure}[t]
\centering
\includegraphics[width=\columnwidth]{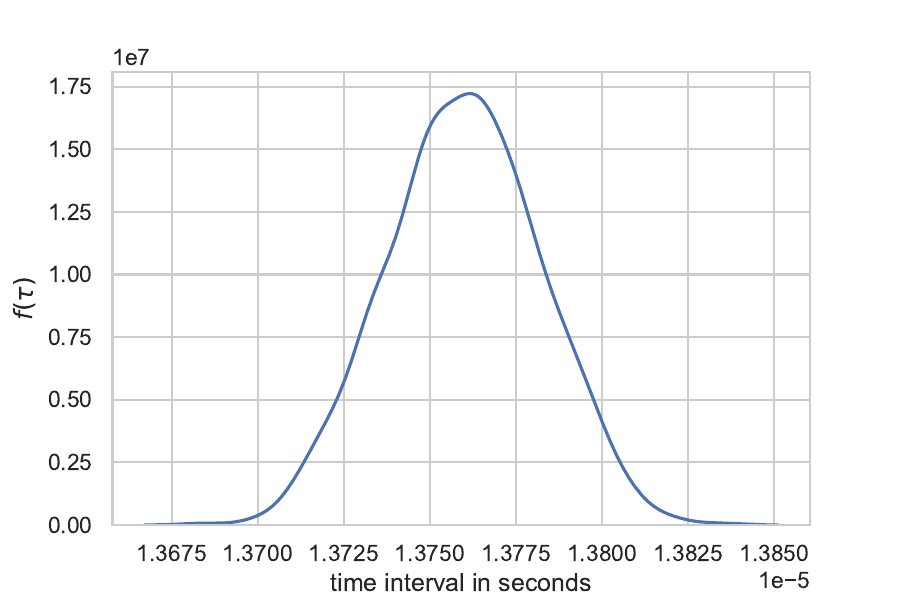}
\caption{Arrival time distribution of benzene in an accelerated field where time is measured in seconds, $V_0=65.96$ V, $V_1=-65.93$ V, $V_2=-1000$ V, and length of the drift region is $L = 0.632$ m.}
\label{fig3}
\end{figure}

\begin{figure}[t]
\centering
\includegraphics[width=\columnwidth]{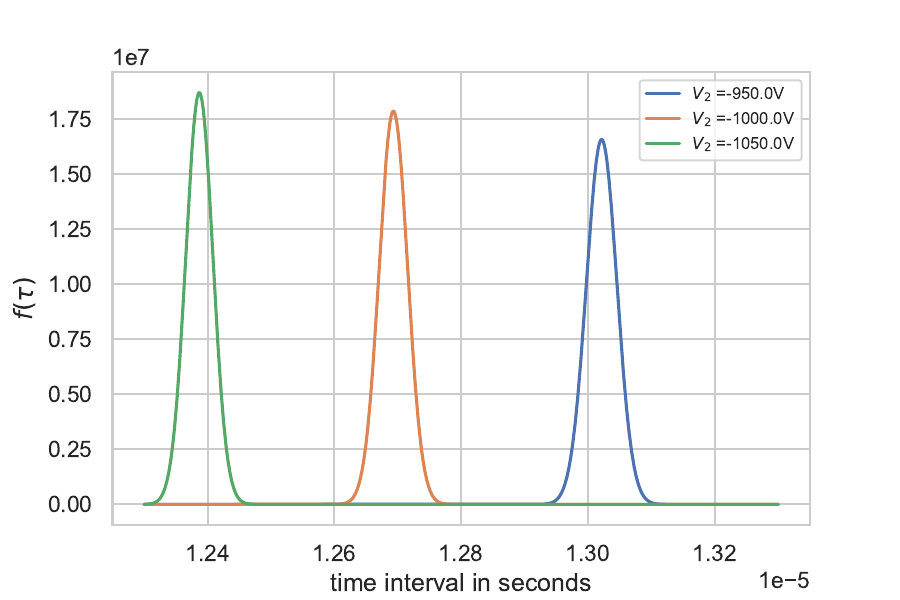}
\caption{Arrival time distribution of benzene with different secondary voltage when $V_0$, $V_1$, and $L$ are constant.}
\label{fig4}
\end{figure}

\begin{figure}[t]
\centering
\includegraphics[width=\columnwidth]{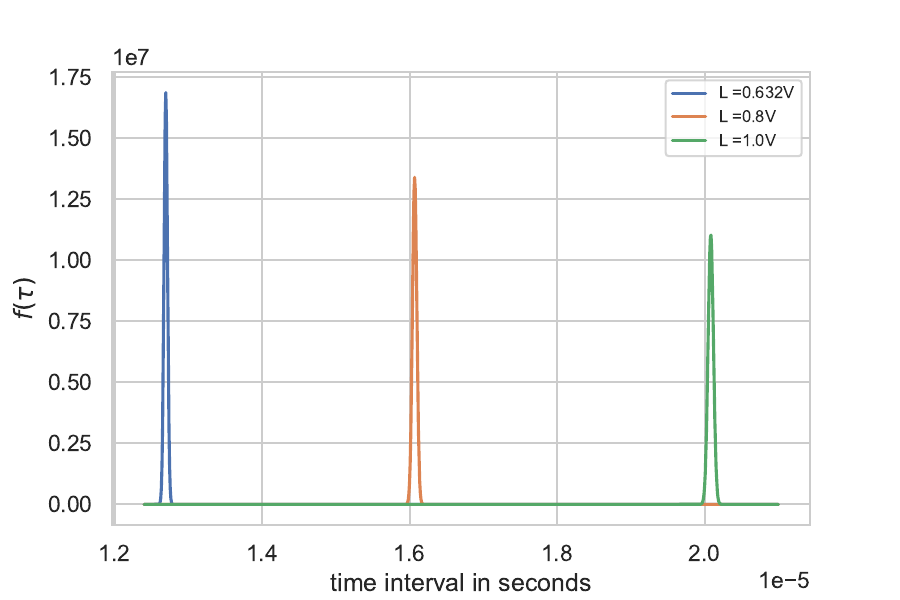}
\caption{Arrival time distribution of Benzene when grid voltages are same but the length of drift region varies.}
\label{fig5}
\end{figure}

The benefit of molecular communication in a vacuum is clearly seen when ions are subjected to an external field source; in this section, we illustrate the effects of external fields on arrival time distributions $f(\tau)$. 

We consider an ionized Benzene molecule having a mass of $78$ atomic mass units (amu) as this is used in \cite{ResolutioninlinearTOF}.%
\footnote{
    The amu is defined as $\frac{1}{12}$ the mass of a carbon-12 atom: $1$ amu $\approx 1.66\cdot 10^{-27}$ kg, approximately the mass of a proton.
} 
As an example, we first consider the parameters used in \cite{ResolutioninlinearTOF}. In this example, the voltages on each grid are $V_0=65.96$ V (Grid $G0$), $V_1=-65.93$ V (Grid $G1$) and $V_2 = -1000$ V (Grid $G2$); the length of the drift region $L$ is $623.6$ mm; and the distances $d_1$ and $d_2$ are $11.6$ mm and $10.0$ mm, respectively. 
Illustrating the arrival time using these parameters in Fig. \ref{fig3}, we can see that the probability density function of Benzene with $78$ amu ranges from $1.3758$ $\mu$s to $1.3763$ $\mu$s while an average number of ions reach around $1.3760$ $\mu$s. The variation in arrival time is thus on the order of nanoseconds.

Continuing the example with benzene, we obtained the arrival time distribution for various voltages $V$ Fig. \ref{fig4}. and lengths of the drift region $L$ Fig. \ref{fig5}. There is a tremendous decrease in average time as the voltage at $G2$ increases. From (\ref{eqn:drifttime}), as voltage increases, ions get more kinetic energy and reach the destination more quickly; moreover, the dispersion of the first arrival time distribution is reduced. 
Meanwhile, in Fig. \ref{fig5}., we see that the average arrival time increases with the increase in length of the drift region and that the dispersion of the first arrival time is also increased.
\begin{table}[h!]
\centering
\begin{tabular}{  m{14em}  } 
 \hline
  Parameters used in the simulation of communication systems  \\
  \hline
  $V_0 = 65.96$V\\
  $V_1 = -65.96$V\\
  $V_2 =$ $-950$, $-1000$, $-1050$V\\\\
  $T= 300K$\\\\
  $L = $ $0.632$m, $0.8$m, $1$m\\\\
  Hydrogen mass = $1.007$ amu\\
  Nitrogen mass = $14.006$ amu\\
  Argon mass = $39.94$ amu\\
  Benzene mass = $78$ amu\\
  \hline 
  \end{tabular}\\
\caption{Parameters of vacuum communication systems}
    \label{tab:Table I}
\end{table}
\subsection{Data Rates} 
\begin{figure}[t!]
\centering
\includegraphics[width=\columnwidth]{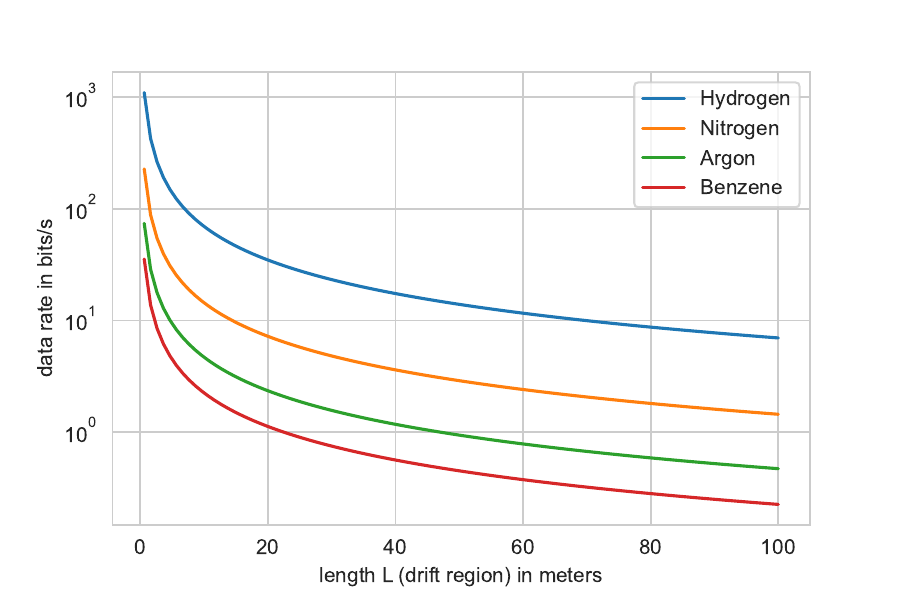}
\caption{Estimated data rate when there is no external field. Ions are generated from elements through some ionization source and are free to move without any electric field source. }
\label{figg}
\end{figure}

Recall our simplified definition of data rate from (\ref{eqn:DataRate}). First, as a point of comparison, we analyzed the system when there was no external field to accelerate ions (i.e. $V_0 = V_1 = V_2 = 0$) so that the initial velocity of the ions is given by their thermal energy. Data rates for hydrogen, nitrogen, argon and benzene are shown in Fig. \ref{figg}., with respect to the length of the drift region. Even without an external accelerating field, information rates in the range of length(0-20 m) are happened to be in between 10-100 bits/s, already 1 or 2 orders of magnitude higher than most results for existing diffusion-based systems.

\begin{figure}[t!]
\centering
\includegraphics[width=\columnwidth]{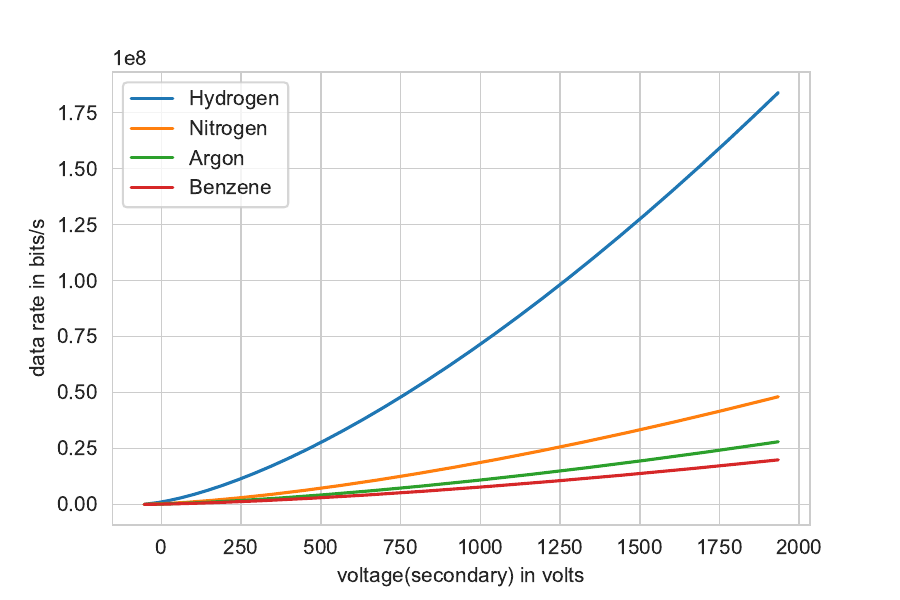}
\caption{Estimated data rate with range of $V_2$ (i.e. voltage on Grid 2). The data rate increases because as the voltage becomes higher, ions get more energy and reach relatively earlier with less dispersion.}
\label{fig8}
\end{figure}

\begin{figure}[t]
\centering
\includegraphics[width=\columnwidth]{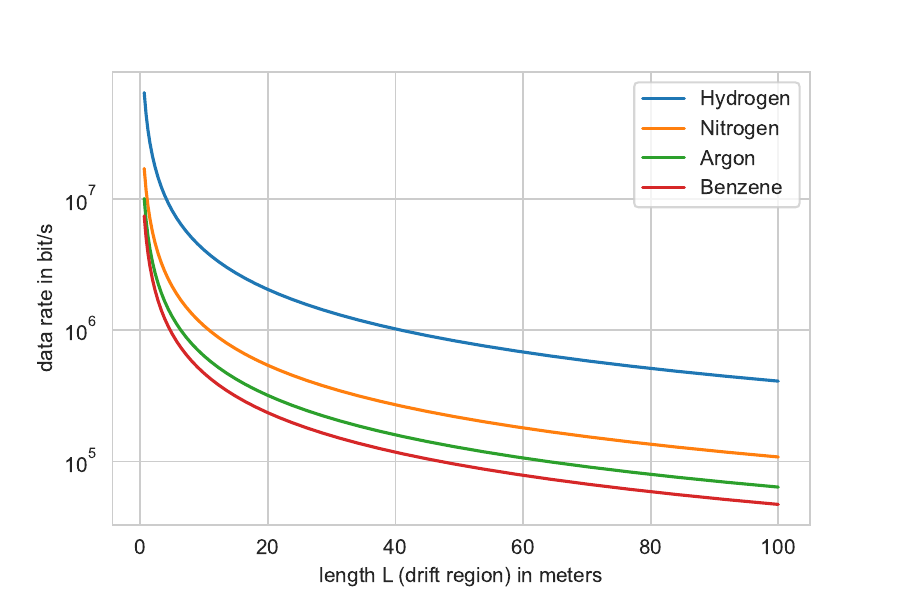}
\caption{Estimated data rate with different length $L$ of the drift region. The data rate decreases as the length of the drift region increases. }
\label{fig9}
\end{figure}

Applying an external field increases the achievable information rate by many orders of magnitude. As one example, for the benzene arrival time distribution in Fig. \ref{fig3}, we have \rev{$T = 1.4 \times 10^{-7}$ s}, which corresponds to the data rate of $7.33$ Mbits/s. Fig. \ref{fig8}. shows estimated data rates for hydrogen, nitrogen, argon and benzene for different secondary grid voltages. Velocity is inversely proportional to mass, so hydrogen, the lightest element, has a very high rate, exhibiting data rates in the hundreds of megabits for high values of $V_2$; even for heavier ions, data rates in the tens of megabits are possible. Fig. \ref{fig9} shows the variation of rate with respect to the length of the drift region. As the drift region's length increases, the arrival time distribution dispersion increases, decreasing the achievable data rate. Nonetheless, over  distances of up to 100 m, data rates in the $10^5$ to $10^6$ bits/s range are possible for most of the considered ions.
It is important to note that these information rates are obtained for {\em single ions}; thus, the information rates may be interpreted as bits/s/molecule. From Fig. \ref{fig8}, hydrogen ions can thus achieve in excess of $100 $ bits/s/molecule.





\subsection{Improvement of data rates}

We modified our design by inserting a velocity selector (Wien filter) to improve data rates. We consider the same physical parameters used in \cite{waahlin1964colutron}: length $b$ of electrostatic plates is $39.37$ cm, and the distance between the aperture and centre of a filter $l$ is $345$ cm. Simulations have been conducted on different aperture sizes. Other parameters are set according to our basic TOFMS setup. Accelerating voltages (i.e. grid voltages) are the same as that in the previous section, while the electric field is set as $4$ kV/m. \rev{We will determine data rates first according to (\ref{eqn:DataRate}) and then compare it with rates obtained from (\ref{eqn:infotheoretic rates}).}

\begin{figure}[t]
\centering
\includegraphics[width=\columnwidth]{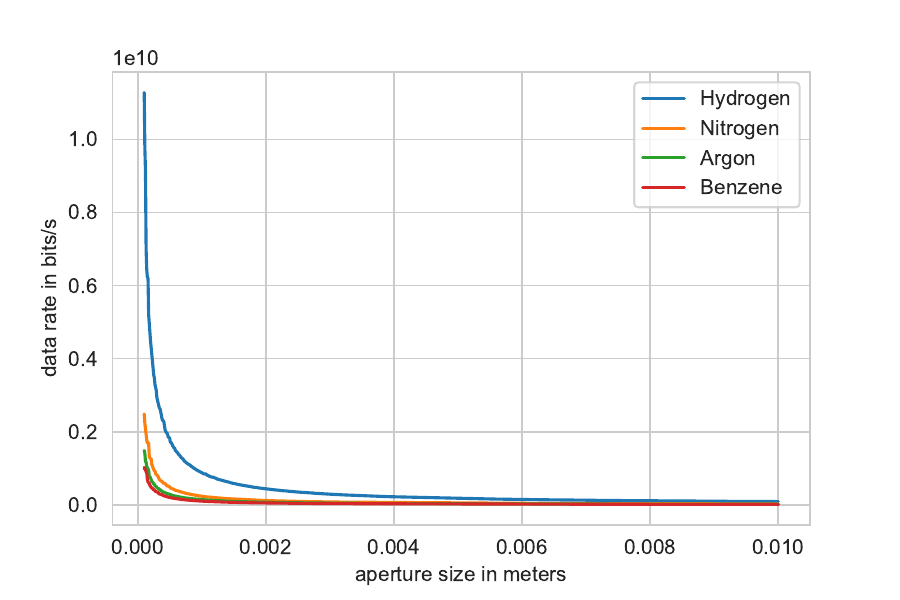}
\caption{Data rate of different elements varies with aperture size. The only source of uncertainty here is the velocity range that can pass through the aperture. As the size of the aperture increases, the rate decreases. Since it allows more velocity components to pass through.}
\label{fig10}
\end{figure}

The Wien filter parameters are chosen to select each element's {\em mean velocity}, which is our information carrier. Fig. \ref{fig10} shows data rates of elements, which are dramatically higher than our simple system without a velocity filter. As the lightest element, hydrogen also has the fastest rate, almost more than $10^9$ bits/s.\par
\begin{figure}[t]
\centering
\includegraphics[width=\columnwidth]{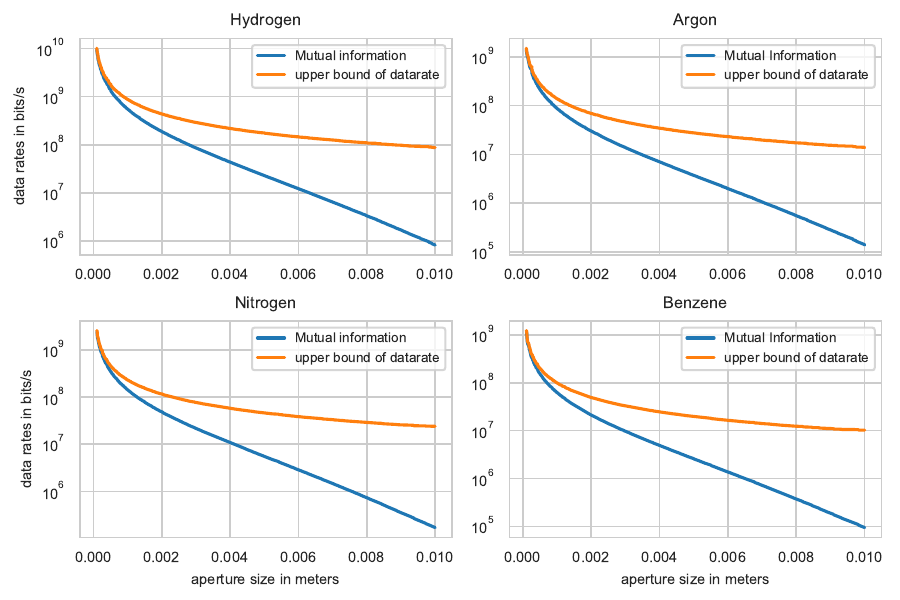}
\caption{Upper bound of data rate has been calculated by using (\ref{eqn:DataRate}) while (\ref{eqn:infotheoretic rates}) is used to determine mutual information. It has added uncertainty of whether or not the selected velocity component is present, resulting in lower mutual information than the upper bound of the data rate. }
\label{fig11}
\end{figure}

 Using an information theoretic approach, we determined the rates by considering if the selected velocity is present in the ionic beam. The effects of the uncertainty are captured by mutual information (\ref{eqn:infotheoretic rates}), assuming that probabilities of sending one and zeros are equal, that is, $P(x=1)=P(x=0)$. Fig. \ref{fig11} clearly shows that the mutual information of all four elements is less than that of our estimated data rates. Data rates have increased significantly with a velocity filter; however, if multiple molecules are sent simultaneously using a molecular MIMO technique, data rates may increase even further.  

\section{Conclusion and future work}
This work has presented an idea to achieve high data rates in molecular communications. A simple two-step TOFMS was considered to demonstrate the concept of a communication system consisting of the transmitter (ion source) and receiver (detector) connected by a vacuum, with ions as information carriers. Data rates achieved by this simple method show that the proposed scheme can be considered  a potential candidate for high-speed molecular communication systems.
  
Given the preliminary nature of the work, many potential areas exist to explore. For example, it would be useful to consider a velocity filter in design to decrease the width of the arrival time distribution and improve the information rate further. Furthermore, different isotopes of an element can also be used to encode information: isotopes being different in mass tend to follow different paths under the influence of a magnetic field. Moreover, information-theoretic analysis of the channel and communication system may be performed to optimize the information rates. \par

These findings can be further explored in terms of applications such as satellite-to-satellite communication. Generally, satellite does not communicate directly with each other. However, the direct communication between satellites may be energy efficient as shown in \cite{Inscribedmatter} where the idea of communication in space is presented via inscribed matter. The channel is already near to the vacuum, so fast communication is possible using ions in space communication. The feasibility of this application can be determined by exploring the further sources of interference/noise and synchronization of the transmitter and receiver. Moreover, this can be investigated in terms of mobile source and destination as satellites may be moving.

\bibliographystyle{ieeetr}
\bibliography{globecom21}

\begin{IEEEbiography}[{\includegraphics[width=1in,height=1.25in,clip]{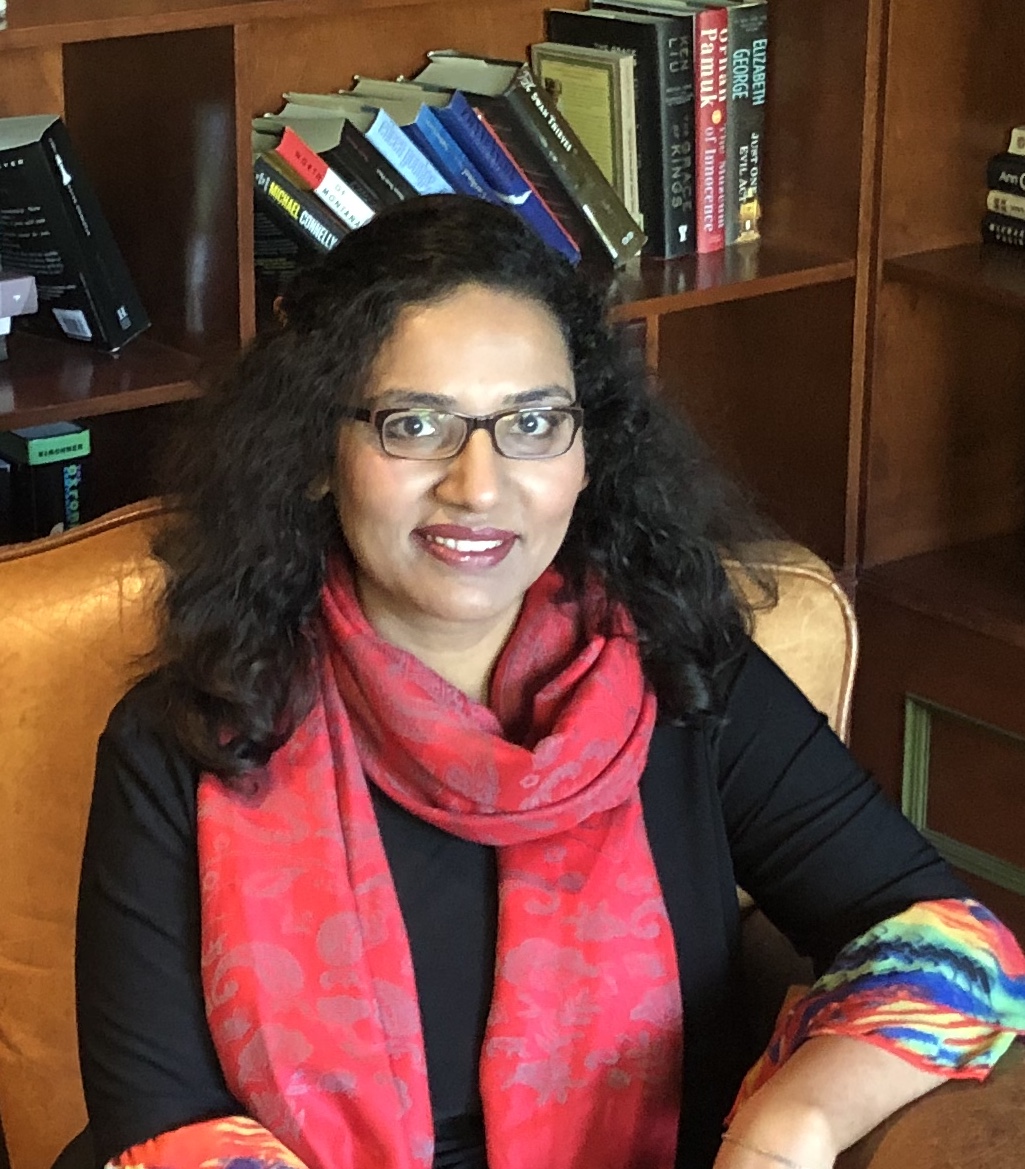}}]{Taha Sajjad} (Student Member, IEEE) is a graduate (Ph.D.) student in Lassonde Electrical and Computer Science
department York University. She is currently working in Eckford’s Lab on the application of
molecular communication. Taha did her M.A.Sc. in electrical engineering from
University of Engineering and Technology (UET), Pakistan, in 2007. Her research has revolved
around multidisciplinary areas, such as communications, embedded systems and biomedical
instrumentation. Taha has also translated a comprehensive book on mind sciences, which
explains how mind power could change one’s life. Moreover, she has been involved in various volunteer activities in the graduate student association and Canadian Science and Policy Centre(CSPC).
\end{IEEEbiography}

\begin{IEEEbiography}[{\includegraphics[width=1in,height=1.25in,clip]{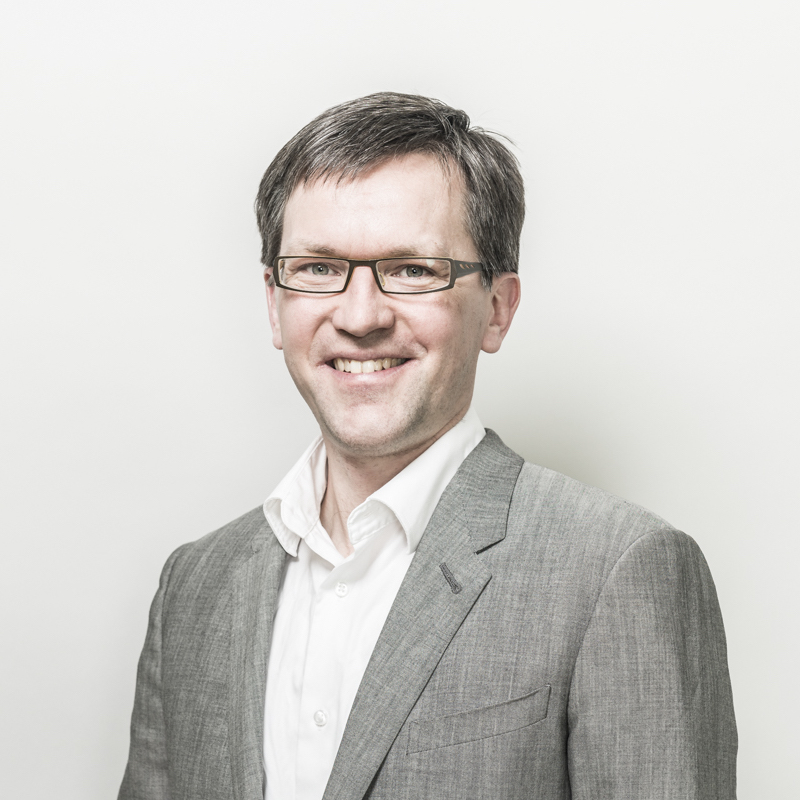}}]{Andrew W. Eckford} (Senior Member, IEEE) received the B.Eng. degree in electrical engineering from the Royal Military College of Canada in 1996, and the M.A.Sc. and Ph.D. degrees in electrical engineering from the University of Toronto in 1999 and 2004, respectively. He was a Postdoctoral Fellowship with the University of Notre Dame and the University of Toronto, prior to taking up a faculty position with York, in 2006. He is an Associate Professor with the Department of Electrical Engineering and Computer Science, York University, Toronto, ON, Canada. He has held courtesy appointments with the University of Toronto and Case Western Reserve University. In 2018, he was named a Senior Fellow of Massey College, Toronto. He is also a coauthor of the textbook Molecular Communication (Cambridge University Press). His research interests include the application of information theory to biology and the design of communication systems using molecular and biological techniques. His research has been covered in media, including The Economist, The Wall Street Journal, and IEEE Spectrum. His research received the 2015 IET Communications Innovation Award, and was a Finalist for the 2014 Bell Labs Prize.
\end{IEEEbiography}

\end{document}